\documentclass{PoS}
\usepackage{bm,amsmath,amssymb}
\usepackage[mathscr]{eucal}

\title{Jet quenching parameter in an expanding QCD plasma}

\ShortTitle{$\hat{q}$ in an expanding QCD plasma}

\author{Edmond Iancu\\
        Institut de Physique Th\'{e}orique, CEA Saclay, 91191, Gif-sur-Yvette Cedex, France
        }
\author{Pieter Taels\\        
        Institut de Physique Th\'{e}orique, CEA Saclay, 91191, Gif-sur-Yvette Cedex, France\\
        Department of Physics, University of Antwerp, Groenenborgerlaan 171, 2020 Antwerpen, Belgium\\
	INFN Sezione di Pavia, via Bassi 6, 27100 Pavia, Italy
}
        
\author{\speaker{Bin Wu}\thanks{https://indico.cern.ch/event/634426/contributions/3090497/attachments/1726178/2788443/hp2018.pdf}\\
        Theoretical Physics Department, CERN, CH-1211 Gen\`eve 23, Switzerland\\
        E-mail: \email{b.wu@cern.ch}}

\abstract{We present a new definition of the jet quenching parameter $\hat{q}$ in a weakly-coupled quark-gluon plasma undergoing boost-invariant longitudinal expansion. We propose a boost-invariant definition of $\hat{q}$, which is proportional to the broadening of the angular variables $\eta$ (the pseudo-rapidity) and $\phi$ (the azimuthal angle). We furthermore consider radiative corrections to $\hat{q}$ and find potentially large corrections enhanced by a double logarithm like the case of a static medium. But unlike for the static medium, these corrections are now local in (proper) time.}

\FullConference{International Conference on Hard and Electromagnetic Probes of High-Energy Nuclear Collisions\\
		30 September - 5 October 2018\\
		Aix-Les-Bains, Savoie, France}

\begin{document}

\section{Introduction}

The transport coefficient $\hat{q}$ is one of the most important parameters for studying jet quenching in relativistic heavy-ion collisions. $\hat{q}$ can be defined as the typical value of the transverse momentum broadening $p_\perp^2$ of a high-energy parton averaged over its path length in QCD matter. It has been shown that medium-induced radiative energy loss is also proportional to $\hat{q}$ \cite{Baier:1996sk}.

Radiative corrections to $\hat{q}$ have been calculated in a static QCD medium with a fixed length $L$. The transverse momentum broadening is found to receive a double logarithmic correction \cite{Wu:2011kc}, which can be resummed in perturbative QCD \cite{Liou:2013qya}. Such a double logarithmic enhancement has been shown to be universal both in $p_\perp^2-$broadening  \cite{Wu:2011kc,Liou:2013qya} and radiative energy loss \cite{Blaizot:2014bha, Wu:2014nca}. And it can be formally absorbed into a renormalized $\hat{q}$ \cite{Blaizot:2014bha, Iancu:2014kga}. 

It is important to measure $\hat{q}$ as transverse momentum broadening. By comparing with the measurement from parton energy loss \cite{Armesto:2005mz, Burke:2013yra}, one can hence test the fundamental relation between $p_\perp^2$ and parton energy loss. Such measurements have been proposed in  \cite{Wu:2010ze, Mueller:2016gko,Chen:2016vem, Chen:2018fqu} (see also a recent measurement by the STAR collaboration at RHIC  \cite{Adamczyk:2017yhe}). 

\begin{figure}
\begin{center}
 \includegraphics[width=0.55\textwidth]{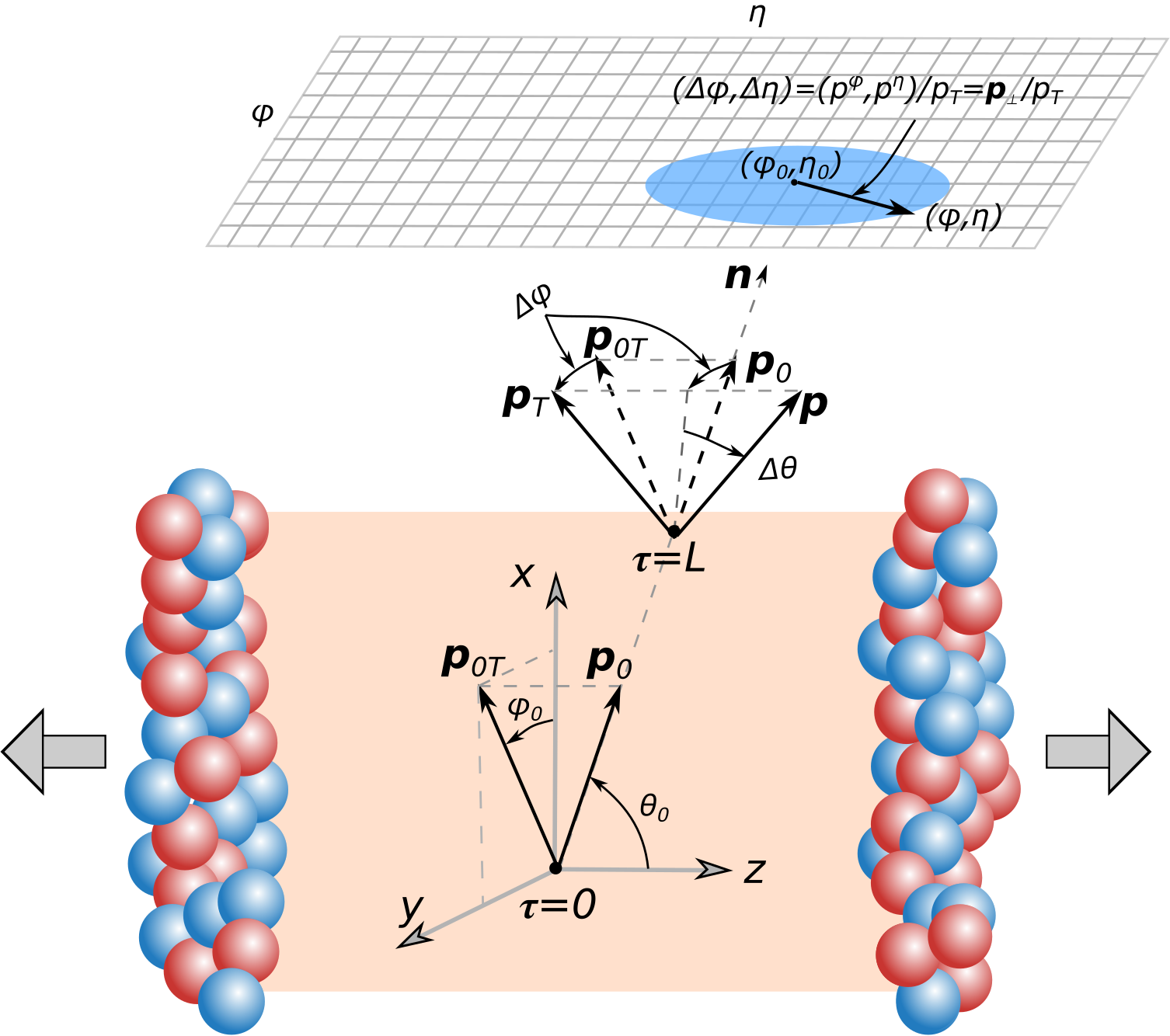} 
 \end{center}
 \caption{A pictorial representation of the definition for transverse momentum broadening in a nucleus-nucleus collision. An energetic parton is created at (proper) time $\tau$=0 with the 4-momentum $p^\mu_0$ and decouples from the medium at $\tau= L$, with the 4-momentum $p^\mu$. The broadening refers to a change $\Delta \phi$ in the azimuthal angle and a change $\Delta \eta$ in the pseudo-rapidity (corresponding to a change $\Delta\theta$ in the polar angle).These changes are summarized in a vector ${\bm {p}}_\perp$ in the 2-dimensional rapidity-azimuthal angle plane $(\Delta\eta,\Delta\phi)$, which is precisely the plane generally used to represent the kinematics of a high-energy collision.}
 \label{fig1}
\end{figure}

As illustrated in Fig. \ref{fig1}, we found that it is convenient to choose a new coordinate system in which $\hat{q}$ can be defined as the broadening of the angular variables $\eta$ (the pseudo-rapidity) and $\phi$ (the azimuthal angle)  \cite{Iancu:2018trm}. This was motivated by the proposal to measure $\hat{q}$ from the broadening of jets in $\Delta\phi$ \cite{Mueller:2016gko,Chen:2016vem, Chen:2018fqu}. We also found that using these new coordinates radiative corrections to $\hat{q}$ in a longitudinally boost-invariant plasma can be calculated in a way similar to the case in the static medium.

\section{$\hat{q}$ as jet broadening in $\Delta\phi$ and $\Delta\eta$}

Let us study the motion of an energetic parton which propagates through a longitudinally-expanding plasma (cf. Fig. \ref{fig1}). We choose the $z$-axis as the collision (or ``longitudinal'') axis and use the notation $\bm{x}_T=(x,y)$ for the coordinates of a point in the transverse plane. The collision starts at $z = t = 0$ and a high-$p_T$ parton, initiating the jet, is produced almost instantaneously at $t=z=0$. It is convenient to use the momentum rapidity\footnote{We treat all particles as massless, so their momentum rapidity is the same as their pseudo-rapidity: $\eta=-\ln\tan(\theta/2)$, with $\theta$ their polar angle: $v_z=\cos\theta$.} $\eta$ and the space-time rapidity $\eta_s$, defined respectively as
\begin{eqnarray}
\eta\equiv\frac{1}{2}\ln\left(\frac{E+p_z}{E-p_z}\right) ,\qquad \eta_s\equiv \frac{1}{2}\ln\left(\frac{t+z}{t-z}\right),
\end{eqnarray}
as well as the azimuthal angle $\phi$, which is measured from the $x$-axis around the beam. In the situation at hand, one can identify the two rapidities, i.e., $\eta=\eta_s$.

In nuclear collisions a QCD jet is usually defined as a cluster of final-state particles in the $\phi$-$\eta$ space. It is hence important to find a convenient way to calculate the broadening of the energetic parton in $(\eta,\phi)$. Let us take $\phi=\phi_0$, $\eta=\eta_0$ and $p^\mu=p_0^\mu=p_T (\cosh\eta_0, \cos\phi_0,\sin\phi_0,\sinh\eta_0)$ at $t=0$. At a later time $\tau\equiv \sqrt{t^2-z^2} > \tau_0$, when bulk matter gets formed, this parton is subject to multiple scattering. Here, we shall focus on the calculation of jet broadening in $\Delta\phi=(\phi-\phi_0)$ and $\Delta\eta=(\eta-\eta_0)$ resulting from multiple scattering.

We shall perform our calculation in a new coordinate system by choosing four independent basis vectors. First, we choose the initial jet direction
\begin{eqnarray}
\label{hat}
\hat{n}^\mu=\frac{p_0^\mu}{p_T}=(\cosh\eta_0, \cos\phi_0,\sin\phi_0,\sinh\eta_0)
\end{eqnarray}
as one basis vector. Second, it is natural to choose two more basis vectors as follows
\begin{eqnarray}
\label{eq:phietaaxes}
\hat\phi^\mu\equiv (0,\sin\phi_0, -\cos\phi_0, 0),\qquad
\hat\eta^\mu\equiv (-\sinh\eta_0, 0, 0, -\cosh\eta_0),
\end{eqnarray}
which span the 2-dimensional vector space encoding the jet broadening in $\Delta\phi$ and $\Delta\eta$. These vectors are orthogonal in the 4-dimensional sense to the initial jet direction: $\hat\eta \cdot \hat{n} =\hat\phi\cdot \hat{n}=0$. For that reason, the respective components $p^\eta \equiv  \hat{\eta}\cdot p = p_{T}\Delta\eta$ and $p^\phi\equiv \hat{\phi}\cdot p = p_{T}\Delta\phi$ will be referred to as  ``transverse'' and collectively denoted with the subscript $\perp$ :  ${\bm p}_\perp\equiv(p^\phi,p^\eta)$. This 2-dimensional vector should not be confused with the {\it other} transverse momentum in the problem, namely $\bm{p}_T=(p_x,p_y) $, which is orthogonal to the collision axis. The fourth basis vector can be chosen to be the norm to the hyper-surface of the constant proper time $\tau$, which takes the form
\begin{eqnarray}
\label{hat}
{\hat{\tau}}^\mu=\left(\cosh\eta_0, 0 , 0,\sinh\eta_0\right).
\end{eqnarray}

In this new coordinate system the "time" is the proper time $\tau$, the "energy" is $p_T$ and the "transverse" momentum broadening is the broadening in $(\eta,\phi)$. The reason is as follows. In terms of these four vectors, one can easily see that
\begin{align}
g^{\mu\nu}=\hat{n}^\mu \hat{\tau}^\nu+\hat{\tau}^\mu \hat{n}^\nu-\hat{n}^\mu \hat{n}^\nu-\hat{\phi}^\mu \hat{\phi}^\nu-\hat{\eta}^\mu \hat{\eta}^\nu,
\end{align}
and, accordingly,
\begin{align}\label{pmuboost}
p^\mu%&=p_{T}\big(\!\cosh(\eta_0+\Delta\eta),\,\cos(\phi_0+\Delta\phi),\,\sin(\phi_0+\Delta\phi),\,\sinh(\eta_0+\Delta\eta)\big)\nn
&\simeq p_T \hat{n}^\mu- p_{T}\Delta\eta\,\hat\eta^\mu - p_{T}\Delta\phi\,\hat\phi^\mu\, \equiv (p_T,0,\Delta\eta, \Delta\phi),\notag\\
x^\mu%&=p_{T}\big(\!\cosh(\eta_0+\Delta\eta),\,\cos(\phi_0+\Delta\phi),\,\sin(\phi_0+\Delta\phi),\,\sinh(\eta_0+\Delta\eta)\big)\nn
&\simeq \tau \hat{\tau}^\mu- x^\eta\hat\eta^\mu - x^\phi \hat\phi^\mu\,\equiv (0,\tau,x^\eta,x^\phi).
\end{align}

All the discussions in the static case can be straightforwardly generalized into the expanding case using such new coordinates. For example, $\hat{q}$ can be defined as
\begin{align}
\label{eq:S0}
\frac{\partial}{\partial \tau} S(x,{\bm r}_\perp)=-\frac{1}{4}\hat{q}_0(x)r_\perp^2 S(x,{\bm r}_\perp),
\end{align}
which looks the same as the corresponding equation in the static medium with the replacement discussed above. The solution to this equation gives the distribution of the parton in the $(\eta,\phi)$ space at $\tau=L$
\begin{align}
 \label{ptGauss}
 \frac{ d N}{d^2\bm{p}_\perp}=&\int \frac{d^2\bm{r}}{(2\pi)^2} e^{-i{\bm p}_\perp\cdot {\bm r}_\perp} S(\bm{r}_\perp) \,\simeq\,\frac{1}{\pi \mathcal{Q}_0^2(L)}\,e^{-\frac{\bm{p}_\perp^2}{\mathcal{Q}_0^2(L)}}\notag\\
 &\Longleftrightarrow\quad \frac{d N}{d\Delta\phi\, d\Delta\eta} \,\simeq\,\frac{p_T^2}{\pi \mathcal{Q}_0^2(L)}\,e^{-\frac{p_T^2(\Delta\phi^2+\Delta\eta^2)}{\mathcal{Q}_0^2(L)}}
\, .\end{align}
In a longitudianlly boost-invariant plasma, one has \cite{Baier:1998yf}
\begin{align}
   \mathcal{Q}_0^2(L)=\int_{\tau_0}^L d\tau\hat{q}_0(\tau)\simeq \hat q_0(L)L\frac{1-(\tau_0/L)^{1-\beta}}{1-\beta}
\end{align}
with
\begin{align}
\label{eq:qhat0}
\hat q_0 (\tau)\,\simeq\,\hat q_0 (\tau_0)\left(\frac{T(\tau)}{T_0}\right)^3\,\simeq\,\hat q_0(\tau_0)\left(\frac{\tau_0}{\tau}\right)^\beta.
\end{align}

%In the discussion below, we shall take a quark jet as an example to study the jet broadening in $\Delta\phi$ and $\Delta\eta$. All our discussions also apply to a gluon jet as long as one replaces $C_F=(N_c^2-1)/2N_c$ with $C_A=N_c$. Here, the number of colors $N_c=3$. The distribution in ${\bf p}_J$ can be computed as the Fourier transform of the $S$-matrix for the elastic scattering of a small quark-antiquark {\em color dipole} in ${\bf r}_J$:
%\begin{align}
% \label{ptbroad}
% \frac{d N}{d^2 {\bf p}_J} \,%=\frac{1}{2 p_T}\int \frac{dp^+ dp^-}{(2\pi)^3}\delta(2p^+ p^--p_J^2)|M(p_0\to p)|^2
%=\,  \frac{d N}{p_T^2 d\Delta\phi d\Delta \eta}
%=  \int \frac{d^2{\bf r}_J}{(2\pi)^2}\,e^{-i {\bf p}_J \cdot {\bf r}_J} \,S({\bf r}_J)\,.
% \end{align}
%This ``color dipole'' is merely a mathematical construction: its ``quark leg''  is the physical quark as viewed in the direct amplitude, while the ``antiquark leg''  is the same quark, but viewed in the complex conjugate amplitude.

\section{Radiative correction to $\hat{q}(\tau)$ in an expanding plasma}

It is also straightforward to calculate the radiative corrections to $p_\perp^2$ and $\hat{q}$ using these coordinates in an expanding plasma. With $\hat{q}_0$ given by Eq. (\ref{eq:qhat0}), the radiative correction to $p_\perp^2$ is given by
\begin{align}
 \label{deltap}
\delta\langle p_\perp^2\rangle = -\nabla_{\bm{r}}^2 \delta S(\bm{r})\Big |_{r_\perp=0}\,,
\end{align}
where the radiative correction from one gluon emission to $S(\bm{r}_\perp)$
\begin{align}
\label{eq:dS}
\delta S({\bm r})&= - \frac{\alpha_s N_c r^2}{2}
~\text{Re}~\int \frac{d\omega}{ \omega^3}  \int_{\tau_0}^{L} d\tau_2 \int_{\tau_0}^{\tau_2} d\tau_1\nonumber\\
&\times \left\{
e^{-\frac{{\bm r}^2}{4}\int_{\tau_0}^{L} d\tau'\hat{q}_0(\tau')[\theta(\tau_1-\tau')+\theta(\tau'-\tau_2)] } \nabla_{{\bm B}_2} \cdot \nabla_{{\bm B}_1} G^{(3)}( {\bm B}_{2}, \tau_2; {\bm B}_{1}, \tau_1;\bm{r})%\right.\nonumber\\
%&-&\left.\left. \nabla_{{\bm B}_2} \cdot \nabla_{{\bm B}_1}  G_0( {\bm B}_{2}, t_2; {\bm B}_{1}, t_1 )
\right\}\Bigg |^{{\bm B}_2 = {\bm r}}_{{\bm B}_2 = 0}\,
\Bigg |^{{\bm B}_1 = {\bm r}}_{{\bm B}_1 = 0},
\end{align}
the propagator for a gluon and a color dipole in the plasma
\begin{align}G^{(3)}( {\bm B}_{2}, \tau_2; {\bm B}_{1}, \tau_1;\bm{r})=G\left({\bm B}_2 - \frac{\bm r}{2},\tau_2;{\bm B}_1 - \frac{\bm r}{2},\tau_1\right),\end{align}
and
\begin{eqnarray}
G({\bm B}_2, \tau_2; {\bm B}_1, \tau_1)\equiv \frac{i \omega}{2\pi D(\tau_2,\tau_1)}\,e^{\frac{-i\omega}{2 D(\tau_2, \tau_1)}[c_1 {\bm B_1^2}+c_2 {\bm B_2^2}-2 {\bm B}_2\cdot {\bm B}_1]},
\end{eqnarray}
with $c_1\equiv c(\tau_2,\tau_1), c_2\equiv c(\tau_1,\tau_2)$ and the following definitions for the functions $D(\tau_2,\tau_1)$ and $c(\tau_2,\tau_1)$ :
\begin{align}\label{Dcdef}
D(\tau_2,\tau_1)&=\pi  \nu  \sqrt{\tau _1 \tau _2}\, \big[J_{\nu }\left(2 \nu  \Omega_1 \tau _1\right) Y_{\nu }\left(2 \nu \Omega_2 \tau _2\right)-J_{\nu }\left(2 \nu  \Omega_2 \tau _2\right) Y_{\nu }\left(2 \nu  \Omega_1 \tau _1\right)\big],\nonumber\\
c(\tau_2,\tau_1)&=\frac{\pi  \nu  \sqrt{\tau _1 \tau _2}\Omega_2}{\sin(\pi  \nu )}\,\big[J_{\nu -1}\left(2 \nu  \Omega_2 \tau _2\right) J_{-\nu }\left(2 \nu  \Omega _1 \tau _1\right)+J_{1-\nu }\left(2 \nu \Omega _2 \tau _2\right) J_{\nu }\left(2 \nu\Omega_1\tau _1\right)\big].
\end{align}
Here, we have used the shorthand notation $\Omega_{1,2}\equiv \Omega(\tau_{1,2})$ with $\Omega(\tau)\equiv \sqrt{i\hat{q}_0(\tau)/\omega}$ and $\nu\equiv 1/(2-\beta)$.

The extraction of the double logarithmic terms in the radiative correction to $p_\perp^2$ from the above equation involves exactly the same manipulations as in the case of the static medium in Ref.~\cite{Liou:2013qya},  except for the replacement  $\hat q_0\to \hat q_0(\tau)$ and for the fact that the integral over the formation time $\tau_{\text{f}}$ should now be restricted to  $\tau_{\text{f}} < \tau$. And we find
 \begin{align}
\label{DLAexp2}
\delta\langle p_\perp^2(L)\rangle
&\,=\,% \frac{2\alpha_s C_F}{\pi} 
\bar{\alpha}\int_{\tau_0}^L d \tau\,\hat q_0(\tau) \int_{\lambda(\tau)}^{\tau}
\frac{d \tau_{\rm f}}{\tau_{\rm f}}
  \int^{\mathcal{Q}_0^2(L)\tau_{\rm f}}_{\hat q_0(\tau)\tau_{\rm f}^2} \frac{d \omega}{\omega} \,\notag\\
  &\,=\,% \frac{2\alpha_s C_F}{\pi} 
\int_{\tau_0}^Ld \tau\,\delta\hat q(\tau),\qquad\mbox{with}\qquad
\delta\hat q(\tau)\,\equiv\,\hat q_0(\tau) \,\frac{\bar{\alpha}}{2} \ln^2\frac{\tau}{\lambda(\tau)}\,.
  \end{align}
Plugging $\hat q_0(\tau)$ in (\ref{eq:qhat0}) and the thermal wavelength 
$\lambda(\tau)=\lambda_0(\tau/\tau_0)^{\beta/3}$, with $\lambda_0\equiv 1/T_0$, into the above equation gives
\begin{align}
\label{deltaQexp}
\delta\langle p_\perp^2(L)\rangle\simeq \begin{cases}
         \displaystyle{\mathcal{Q}_0^2(L)\,\frac
{2\bar{\alpha}}{27}\,\ln^2\frac{L}{\tau_0}} &
       \mbox{for $\beta=1$\,,}
        \\*[0.2cm]
          \displaystyle{
        \mathcal{Q}_0^2(L)\,\frac
{\bar{\alpha}}{2}\left(1-\frac{\beta}{3}\right)^2\ln^2\frac{L}{\tau_0}}& \mbox{for $\beta< 1$\,.}
    \end{cases}
    \end{align}
Here, we have neglected the difference between $\tau_0$ and $\lambda_0$ within the arguments of the various logarithms. 

It is also straightforward to carry out the all-order resummation of the leading double-logarithmic terms
and we obtain
\begin{align}
 \label{qyy}
 \hat{q}(\tau) = \hat{q}_0(\tau)\,
 \frac{I_1\big(2 \sqrt{\bar{\alpha}}\, Y\big)}{\sqrt{\bar{\alpha}}\,Y}
 = \hat{q}_0(\tau)\, 
 \frac{e^{2 \sqrt{\bar{\alpha}}\, Y}}{\sqrt{4\pi}\,(\sqrt{\bar{\alpha}}Y)^{3/2}}
 \left[1 + \mathcal{O}(1/\sqrt{\bar{\alpha}}Y) \right],
 \end{align}
where $Y=\ln(\tau/\lambda(\tau))$ and the second equality holds when $Y \gg 1/\sqrt{\bar{\alpha}}$, i.e. for sufficiently large time. Eq. (\ref{qyy}) is formally similar to the corresponding result for a static medium except that it depends on $\tau$ instead of $L$ via the $\tau$-dependence of the functions $\hat{q}_0(\tau)$ and $\lambda(\tau)$. Unlike the case in the static medium, this makes it possible to treat the renormalized $\hat q(\tau)$ as a {\em (quasi)local} transport coefficient, like its tree-level counterpart $\hat{q}_0(\tau)$.

\end{document}